\documentclass{aa}
\usepackage{psfig,natbib} 
\usepackage{txfonts}
\def\feka{Fe K$\alpha$}
 
\def\xmm{{\it XMM-Newton}}

\def\exosat{{\it EXOSAT}}
\def\rxte{{\it RXTE}}

\def\rosat{{\it ROSAT}}

\def\3c{3C 390.3}
\def\ltsima{$\; \buildrel < \over \sim \;$}
\def\simlt{\lower.5ex\hbox{\ltsima}} 
\def\gtsima{$\; \buildrel > \over \sim \;$}
\def\simgt{\lower.5ex\hbox{\gtsima}} 

\begin{document}
\title{Non-stationary variability in AGN: the case of 3C~390.3}
\author{M. Gliozzi\inst{1}
\and I.E. Papadakis\inst{2}
\and  R.M. Sambruna\inst{1}
\and M. Eracleous\inst{3}}
\offprints{mario@physics.gmu.edu}
\institute{George Mason University, Department of Physics \&
Astronomy \& School of Computational Sciences, 4400 University Drive, 
MS 3F3, Fairfax, VA 22030
\and Physics Department, University of Crete, 710 03 Heraklion,
Crete, Greece
\and  The Pennsylvania State University, Department of
Astronomy \& Astrophysics, 525 Davey Lab, University Park, PA 16802
             }

\date{Received ...; accepted ...}

\abstract{We use data from a two-year intensive RXTE monitoring
  campaign of the broad-line radio galaxy \object{3C~390.3} to
  investigate its stationarity. In order to exploit the potential
  information contained in a time series more efficiently, we use a
  multi-technique approach.  Specifically, the temporal properties are first
  studied with a non-linear technique borrowed from non-linear dynamics.
  Then we utilize traditional linear techniques both in the Fourier
  domain, by estimating the power spectral density, and in the time
  domain with a structure function analysis. Finally we investigate
  directly the probability density function associated with the
  physical process underlying the signal. All the methods demonstrate
  the presence of non-stationarity. The structure function analysis, and (at a 
  somewhat lower significance level) the power spectral density  suggest that
 3C 390.3 is not even second order stationarity.
  This result indicates, for the first time, that the variability
  properties of the active galactic nuclei light curves may also vary
  with time, in the same way as they do in Galactic black holes,
  strengthening the analogy between the X-ray variability properties
  of the two types of object.   }

   \maketitle
%

\section{Introduction}

Among the basic properties characterizing an Active Galactic Nucleus
(AGN) the flux variability, especially at X-rays, is one of the most
common.  Previous multi-wavelength variability studies (e.g., Edelson
et al. 1996; Nandra et al. 1998) have shown that AGN are variable in
every observable wave band, but the X-ray flux exhibits variability of
larger amplitude and on time scales shorter than any other energy
band, indicating that the X-ray emission originates in the innermost
regions of the central engine.

However, even though the discovery of X-ray variability dates back
more than two decades, its origin is still poorly
understood. Nevertheless, it is widely acknowledged that the
variability properties of AGN are an important means of probing the
physical conditions in their emission regions.  The reason is that
using energy spectra alone it is often impossible to discriminate
between competing physical models and thus the complementary
information obtained from the temporal analysis is crucial to break
the spectral degeneracy.

Over the years, several variability models have been proposed,
involving one or a combination of the fundamental components of an
AGN: accretion disk, corona, and relativistic jet. These models can be
divided into two main categories: 1) intrinsically linear models, as
the shot noise model (e.g., Terrel 1972), where the light curve is the
result of the superposition of similar shots or flares produced by
many independent active regions, and 2) non-linear models, which
require some coupling between emitting regions triggering avalanche
effects, as the self-organized criticality model (e.g., Mineshige et
al.1994), or models assuming that the variability is caused by
variations in the accretion rate propagating inwards (e.g. Lyubarskii
1997).  The ``rms-flux relation'' recently discovered by Uttley \&
McHardy (2001), along with detection of non-linearity in different AGN
light curves (i.e., in 3C~345 by Vio et al. 1992; in 3C~390.3 by
Leighly \& O'Brien 1997; in NGC~4051 by Green et al. 1999;
in Ark~564 by Gliozzi et al. 2002), favors
non-linear variability models, but there is no general consensus on
the nature of the X-ray variability yet.

Because of their brightness, the temporal and spectral properties of
Galactic black holes (GBHs) are much better known and can be used to
infer information on their more powerful, extragalactic analogues, the
AGN. For example, it is well established that GBHs undergo state
transitions (see McClintock \& Remillard 2003 for a recent review),
switching between ``low'' and ``high'' states, which are both
unambiguously characterized by a specific combination of energy
spectrum and power spectral density (PSD).

Early AGN PSD studies based on \exosat\ data (e.g., Lawrence \&
Papadakis 1993; Green et al. 1993) revealed a red-noise variability
(i.e., a PSD described by a power law), similar to the one observed in
GBHs, which in addition show breaks in their PSD (i.e., departure from
the power law) at specific frequencies depending on the spectral
state. Recently, power-spectral analysis of high-quality light curves
of a few AGN (obtained with \rxte\ and \xmm\, e.g., Uttley, et
al. 2002; Markowitz et al. 2003; McHardy et al. 2004) 
has led to the first unambiguous
detection of ``frequency breaks'' in their power spectra,
strengthening the similarity between the AGN and GBH PSDs. This
similarity lends further support to the hypothesis that the same
physical process is at work in AGN and GBHs, regardless of the black
hole mass.

Almost all methods employed in timing analysis (e.g., power spectrum,
structure function, auto-correlation analysis) require some kind of
stationarity. Generally speaking, a system is considered stationary,
if the average statistical properties of the time series are constant
with time. More specifically, the stationarity is defined as ``strong''
if all the moments of the probability distribution function are
constant, and ``weak'' if only the moments up to the second order
(i.e., mean and variance) are constant.
Usually, non-stationarity is considered an undesired
complication. However, in some cases the non-stationarity provides the
most interesting information, as in the case of GBHs switching from a
spectral state to another.
\begin{figure*}[th]
\psfig{figure=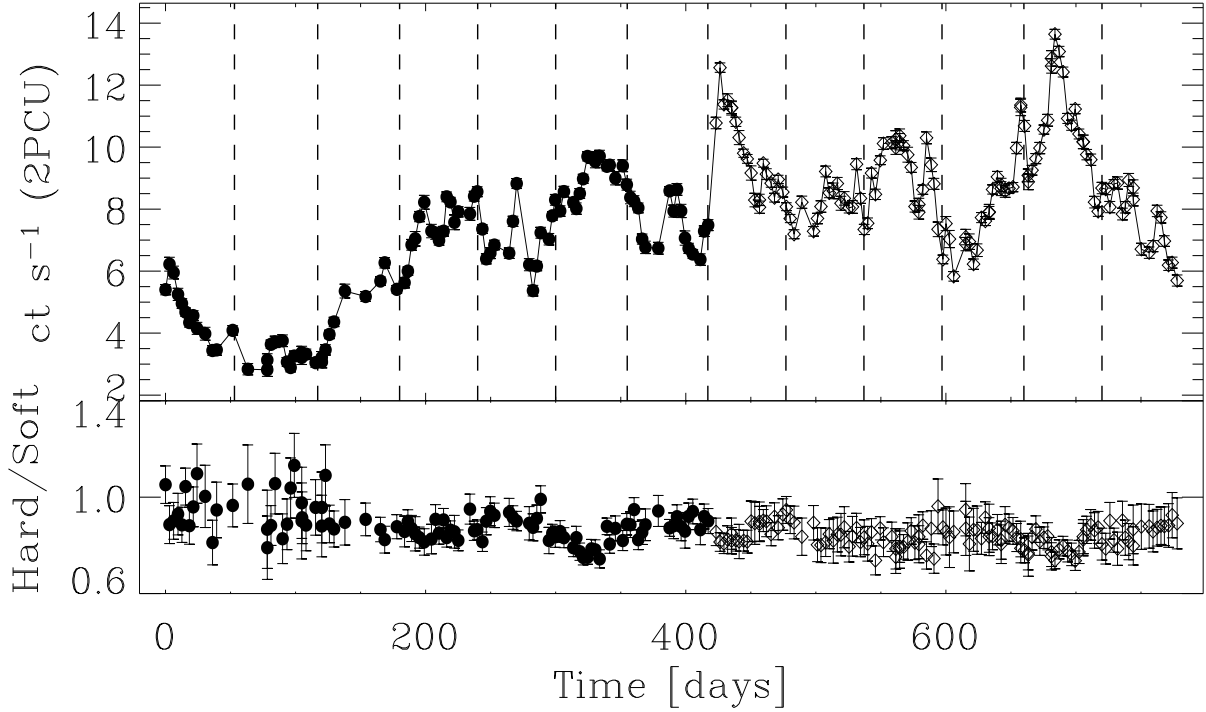,height=10cm,width=16.5cm,%
bbllx=32pt,bblly=6pt,bburx=530pt,bbury=290pt,angle=0,clip=}
\caption{X-ray light curves of the background-subtracted count rate in
the 2--20 keV band (top panel) and of the X-ray color 7--20 keV/2--5
keV (bottom panel) from RXTE PCA observations of 3C 390.3.  Time bins
are 5760s ($\sim$ 1 RXTE orbit). Filled circles represent data points
from the first monitoring campaign from 1999 January 8 to 2000
February 29, whereas open diamonds are data points from the second
monitoring campaign from 2000 March 3 to 2001 February 23. The
dashed lines in the top panel indicate the intervals used for the
time-resolved spectral analysis (see $\S$9).}
\label{figure:lc}
\end{figure*}

Compared to their scaled-down counterparts, the GBHs, the study of
non-stationarity in AGN is complicated by the fact that their flux
variability is characterized by much longer time-scales. This might be
the reason why ``different'' states have not been observed yet in AGN,
even though a correspondence between GBH spectral states and some AGN
classes (namely, between Seyfert 1 galaxies and the GBH low state, and
between Narrow-Line Seyfert 1 galaxies and the GBH high state,
respectively) has been hypothesized.  Nonetheless, several monitoring
campaigns carried out by \rxte\ in the last decade have produced
archival data sets suitable for searching for non-stationarity in the
AGN light curves. Such a detection of non-stationarity would suggest
transitions of the source between different states.

We use data from a two-year intensive \rxte\ monitoring campaign of
the broad-line radio galaxy (BLRG) \object{3C~390.3} to search for
evidence of non-stationarity. Previous X-rays studies with
\rosat, and \rxte\ (Leighly \& O'Brien 1997, Gliozzi et
al. 2003) demonstrated that 3C~390.3 is one of the most variable AGN
on time scales of days and months. In addition, the \rosat\ HRI
monitoring campaign has shown evidence for non-linearity and possibly
non-stationarity (Leighly \& O'Brien 1997). In order to exploit all
the information contained in the time series, we employ a number of
complementary methods. Specifically, we investigate the stationarity
in \object{3C~390.3} not only with traditional linear techniques as
the PSD in the Fourier domain, and the structure function (SF) and
auto-correlation function (ACF) in the time domain, but also in
pseudo-phase space, with techniques borrowed from nonlinear
dynamics, and finally we probe directly the probability density
function (PDF) associated with the physical processes underlying the
signal.

\section{Observations and Data Reduction}
We use archival \rxte\ data (PI: Leighly) of 3C~390.3.  This source
was observed by \rxte\ for two consecutive monitoring campaigns
between 1999 and 2001.  The first set of observations was carried out
from 1999 January 8 to 2000 February 29, and the second one from 2000
March 3 to 2001 February 23.  Both campaigns were performed with
similar sampling: 3C~390.3 was regularly observed for $\sim$
1000--2000 s once every three days.  The observations were carried out
with the Proportional Counter Array (PCA; Jahoda et al. 1996), and the
High-Energy X-Ray Timing Experiment (HEXTE; Rotschild et al. 1998) on
\rxte. Here we will consider only PCA data, because the
signal-to-noise of the HEXTE data is too low for a meaningful timing
analysis.

The PCA data were screened according to the following acceptance
criteria: the satellite was out of the South Atlantic Anomaly (SAA)
for at least 30 minutes, the Earth elevation angle was $\geq 10^{\circ}$,
the offset from the nominal optical position was $\leq
0^{\circ}\!\!.02$, and the parameter ELECTRON-0 was $\leq 0.1$. The
last criterion removes data with high particle background rates in the
Proportional Counter Units (PCUs). The PCA background spectra and
light curves were determined using the ${\rm L}7-240$ model developed
at the \rxte\ Guest Observer Facility (GOF) and implemented by the
program {\tt pcabackest} v.2.1b. Since the two monitoring campaigns
span three different gain epochs, the appropriate files 
provided by the \rxte\ Guest Observer Facility (GOF), were used to
calculate the background light curves. This model is appropriate
for ``faint'' sources, i.e., those producing count rates less than 40
${\rm s^{-1}~PCU^{-1}}$. All the above tasks were carried out using
the {\tt FTOOLS} v.5.1 software package and with the help of the {\tt
  rex} script provided by the \rxte\ GOF, which also produces response
matrices and effective area curves appropriate for the time of the
observation. Data were initially
extracted with 16 s time resolution and subsequently re-binned at
different bin widths depending on the application.  The current
temporal analysis is restricted to PCA, STANDARD-2 mode, 2--20 keV,
Layer 1 data, because that is where the PCA is best calibrated and
most sensitive. Since PCUs 1, 3, and 4 were frequently turned off
(the average number of PCUs is 2.25 and 1.95
in the 1999 and 2000 observations, respectively),
only data from the other two PCUs (0 and 2) were used. All quoted
count rates are for two PCUs.

The spectral analysis of PCA data was performed using the
{\tt XSPEC v.11} software package (Arnaud 1996). We used PCA response
matrices and effective area curves created specifically for the
individual observations by the program {\tt pcarsp}, taking into
account the evolution of the detector properties. 
All the spectra were rebinned so that each bin
contained enough counts for the $\chi^2$ statistic to be valid. Fits
were performed in the energy range 4--20 keV,
where the signal-to-noise ratio is the highest.

\section{The X-ray Light Curve}

Figure~\ref{figure:lc} shows the background-subtracted 2--20 keV count
rate and hardness ratio (7--20 keV/2--5 keV) light curves of 3C~390.3.
Filled circles represent data points from the first monitoring
campaign, whereas open diamonds are used for the second campaign.  A
visual inspection of the light curve suggests the existence of
different variability patterns during the two monitoring campaigns. In
the first one, 3C~390.3 exhibits an initial smooth decrease in the
count rate (a factor $\sim$2 in 100 days) followed by an almost steady
increase (a factor $\sim$3.5 in 100 days). After day $\sim$ 200 (from
the beginning of the observation), the source shows significant,
low-amplitude, fast variations. During the second monitoring campaign,
the source does not show any long-term, smooth change in count rate,
but only a flickering behavior with two prominent flares. This
qualitative difference is supported by the difference of the two mean
count rates averaged over each campaign,
$6.4\pm0.2{~\rm s^{-1}}$ and $9.0\pm0.1{~\rm s^{-1}}$
respectively, where the uncertainties, $\sigma/\sqrt n$, were
calculated assuming that the data distribution is normal and the data
are a collection of independent measurements. However, the data points
of AGN light curves are not independent but correlated and thus the
apparent difference between the mean count rates could simply result
from the red-noise nature of the variability process, combined with
its randomness.

Any AGN light curve is just one of the possible realizations of an
underlying physical process, which can be linear stochastic,
non-linear stochastic, chaotic, or a combination of the three
possibilities.  The distinction between stochastic and chaotic
processes is that the former are random, whereas the latter are
deterministic. A further difference is that chaotic processes are
intrinsically non-linear, while stochastic processes can be either
linear or non-linear (in a mathematical sense linearity means that the
value of the time series at a given time can be written as a linear
combination of the values at previous times plus some random
variable).  As a consequence, light curves produced by the same
process may appear different, but the average statistical properties
should in principle be constant, provided that the system is actually
stationary, and the light curves are longer than the characteristic
time scale of the system. This indicates that only average statistical
properties can provide information on the physics underlying a
variability process, whereas instantaneous properties (such as the
visual appearance of the light curves) may be misleading. With this in
mind, we carry out a number of tests for stationarity, which we
describe in detail in the next section.

\section{Stationarity Tests}

In previous temporal studies of AGN, no strong evidence for
non-stationarity has been reported. Indeed, AGN light curves are
considered to be ``second-order'' stationary, due to their red noise
variability. However, the absence of non-stationarity might be partly
due to the fact that previous works were based only on a PSD analysis,
which may not be the most appropriate method to detect
non-stationarity in some cases, as we demonstrate below.

Here, we adopt several complementary methods to investigate the issue
of the stationarity in the \object{3C~390.3} light curve.  First,
we utilize the scaling index method, a technique borrowed from
non-linear dynamics, useful for discriminating between time
series. Secondly, the temporal properties are studied first with
traditional linear techniques in the Fourier domain, by estimating the
PSD, and in the time domain with structure function and
auto-correlation function analyses. Finally, we probe directly the
probability density function associated with the physical process
underlying the signal. We demonstrate that all methods, with the
exception of the PSD, point very strongly to non-stationarity.

\subsection{Non-linear Analysis: Scaling Index Method}

Non-linear methods are rarely employed in the analysis of AGN light
curves, partly because these methods are less developed than the
linear ones, partly because it is implicitly assumed that AGN light
curves are linear and stochastic. However, non-linear methods can be
useful not only to characterize the nature of chaotic deterministic
systems, but also to discriminate between two time series, regardless
of the fact that they are linear or non-linear. 

The scaling index method (e.g., Atmanspacher et al. 1989) is employed
in a number of different fields because of its ability to discern
underlying structure in noisy data. It was applied to AGN time series
by Gliozzi et al. (2002) to show that the low and high states of the
Narrow-Line Seyfert 1 galaxy \object{Ark~564}, indistinguishable on
the basis of linear methods, are actually intrinsically different.  A
detailed description of the method is given in Gliozzi et al. (2002)
and references therein; here we summarize its basic
characteristics.

When methods of nonlinear dynamics are applied to time series analysis, the concept of phase space reconstruction represents the basis for most of the analysis tools. In fact, for studying chaotic deterministic systems it is 
important to establish a vector space (the phase space) such that specifying a point in this space specifies a state of the system, and vice versa. However,
in order to apply the concept of phase space to a time series, which is a scalar sequence of measurements, one has to convert it into a set of state vectors.

For example, starting from a time series, one can construct a set of
2(3, 4, ...)-dimensional vectors by selecting  pairs (triplets, quadruplets,
...) of data points, whose count rate values represent the values of
the different components of each vector. The data points defining the vector
components are chosen such that
the second point is separated from the first by a time delay $\Delta
t$, while the second is separated from the first by $2 \Delta t$, and so on. This
process, called time-delay reconstruction, can be easily generalized to any
n-dimensional vector.  In non-linear dynamics, where this technique is
used to determine fractal dimensions and discriminate chaos from
stochasticity, the choice of the time delay is not arbitrary, but
should satisfy specific rules (see Kantz \& Schreiber 1997 for a
review). In our case, however, this technique is used only to
discriminate between two states of the system (determining any kind of
dimension would be meaningless for non-deterministic systems). Thus
the only requirement here is that the results are statistically
significant. Nonetheless, we have used different values for $\Delta t$,
and verified that the results do not critically depend upon the choice of
the time delay.

The set of n-dimensional vectors, derived from the original light curve,
are then mapped (or embedded)
into an n-dimensional vector space, producing a phase space 
portrait, whose topological properties represent the
temporal properties of the corresponding time series. 

In order to quantify the difference between two phase space 
portraits, a suitable quantity is the correlation 
integral $C(R)$ (Grassberger 
\& Procaccia 1983), which basically counts the number of pairs of points with 
distances smaller than $R$.  
An exhaustive description of the correlation integral method and its application
to X--ray light curves of AGN is given by Lehto et al (1993). 
\begin{figure}
\psfig{figure=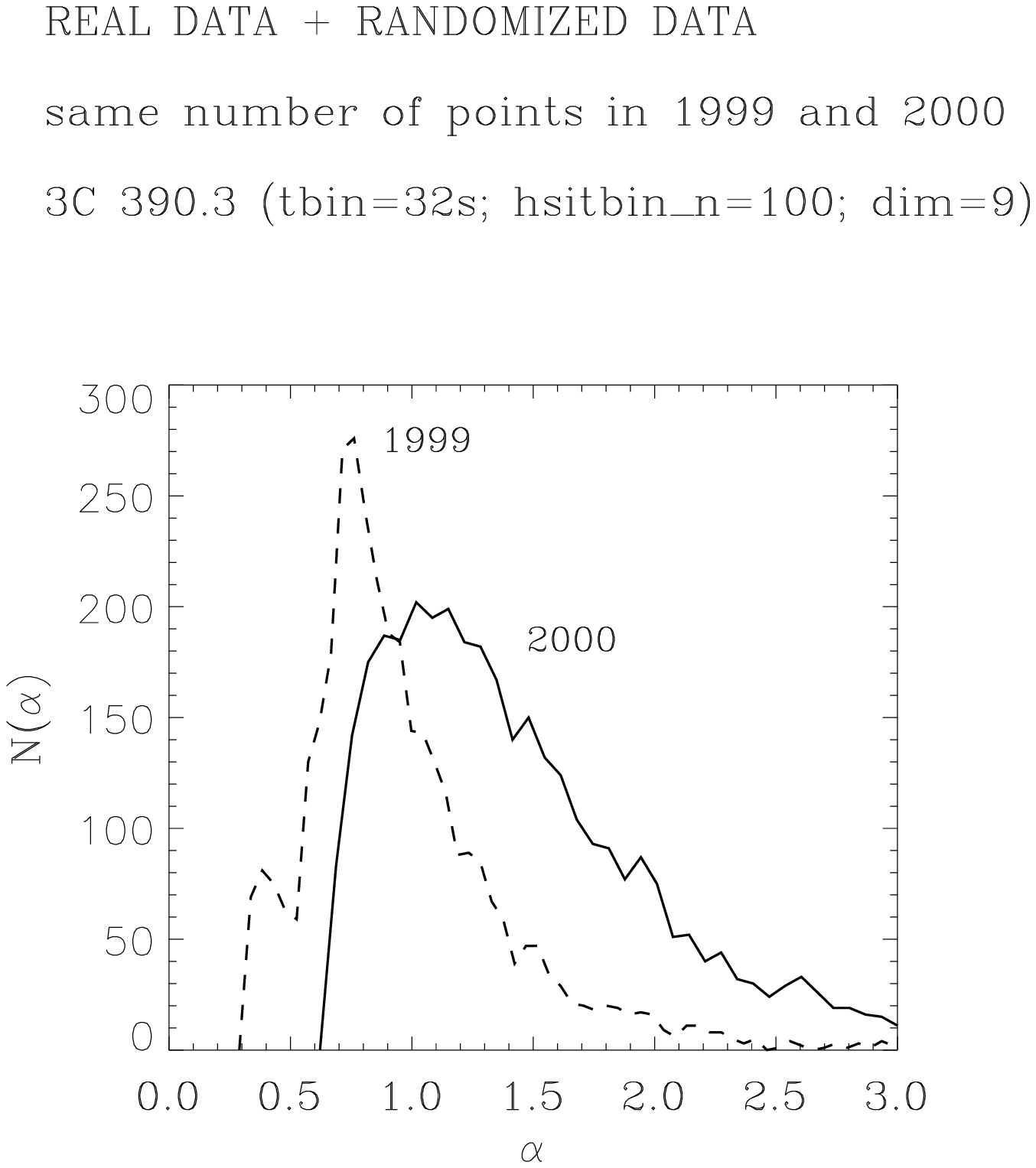,height=8.7cm,width=8.7cm,%
bbllx=47pt,bblly=121pt,bburx=411pt,bbury=434pt,angle=0,clip=}
\caption{Scaling index distributions of \object{3C~390.3} during
the 1999 (dashed line) and 2000 (solid line) monitoring
campaign. 
}
\label{figure:alpha}
\end{figure}

The scaling index method, which is based 
on the local estimate of the correlation integral for each point in the phase 
space,
characterizes quantitatively the data point distribution 
by estimating the ``crowding'' of the data around each data point.
More specifically, for each of the
N points, the cumulative number function is calculated
\begin{equation} 
N_i(R)=n\{j\vert d_{ij}\leq R\},
\end{equation}

In other words, $N_i(R)$ measures the number of points $j$, whose distance 
$d_{ij}$ from a point $i$ is smaller than $R$. 
The function $N_i(R)$, in a given range of radii (which are related 
to the typical distances between the data points, that, in turn, depend on 
the choice of the embedding space dimension), 
can be approximated by a power law
$N_i(R)\sim R^{\alpha_i}$,
where $\alpha_i$ is the scaling index. Explicitly, 
the scaling index for each point $i$ is obtained by 
calculating the cumulative number function $N_i(R)$ at two different
radii, $R_1$ and $R_2$, and by computing the logarithmic slope:

\begin{equation} 
{\alpha_i}=\frac{\log N_i(R_1)-\log N_i(R_2)}{(\log R_1-\log R_2)}.
\end{equation}
Note that for a purely random process the average scaling index  
$\langle\alpha\rangle$ tends to the value of the dimension of the embedding
space, whereas for correlated processes and for deterministic 
(chaotic) processes the value of $\langle\alpha\rangle$ is always smaller 
than the dimension of the embedding space and independent of that dimension.

We have calculated the scaling index for all the points of the two
phase space portraits derived from the two monitoring campaigns, using
embedding spaces of dimensions ranging between 3 and 10.  There are no
systematic prescriptions for the choice of the embedding
dimension. The discriminating power of the statistic based on the
scaling index is enhanced using high embedding dimensions (this can be
understood in the following way: if the data are embedded in a
low-dimension phase-space many of them will fall on the same position,
losing in this way part of the information). On the other hand, the
choice of too high embedding dimensions would reduce the number of
points in the pseudo phase-space, lowering the statistical
significance of the test.  In all cases a significant difference from
the two light curves was found.  The resulting histograms for a
5-dimensional phase space are plotted in Fig~\ref{figure:alpha}: the
dashed line represents data from the first monitoring campaign,
whereas the continuous line indicates points from the second
monitoring campaign. 

There is a remarkable difference between the two distributions, both
in shape and location of the centroid. For this kind of analysis the
most important quantity is the location of the centroid of the
histogram, which is related to the correlation dimension.  Such a
difference cannot be ascribed to the difference in mean count rates
showed during the two monitoring campaigns.  In fact, the scaling
index analysis is sensitive only to the relative positions of the data
points within a distribution, not to their absolute positions in the
phase space. This has been verified applying the scaling index
analysis to randomized light curves, which  were obtained by randomizing
the temporal positions of the data points of each light curve.  In
this way, the statistical properties (i.e., the mean, variance, and
the statistical moments of higher order) of the light curves are
unchanged, whereas the timing structure is destroyed. As expected,
the centroids of the histograms of the randomized distributions
have larger values and are nearly indistinguishable.

We conclude that, the intrinsic, timing properties of the 1999 light
curve are significantly different from the properties of the 2000
light curve. This result suggests that the X-ray emitting process in
3C~390.3 is non stationary.

\subsection{Analysis in the Fourier Domain: Power Spectral Density}
\begin{figure}
\psfig{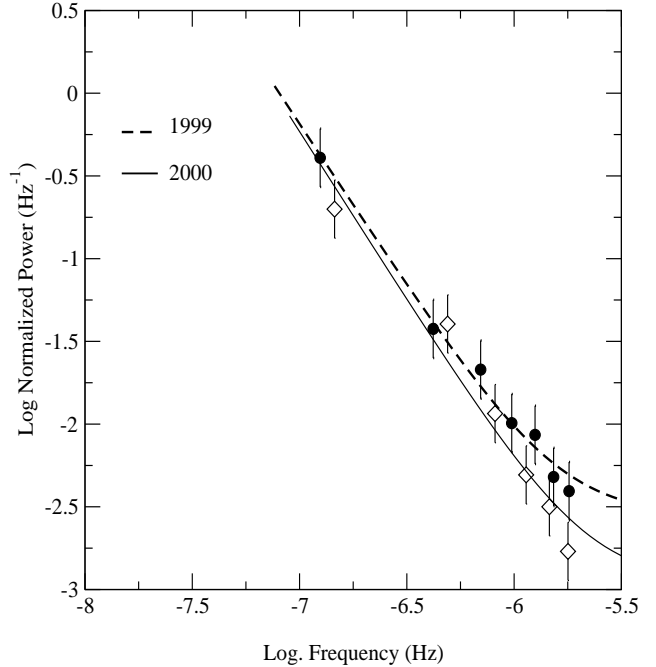}
\caption{RXTE PCA power density spectra using based on the 1999 and
2000 light curves with 3-day bins, normalized to their mean (filled
circles and open diamonds, respectively). The solid and dashed lines
show the best fitting power law models, including the effects of the
experimental Poisson noise level.}
\label{figure:psd}
\end{figure}

We followed the method of Papadakis \& Lawrence
(1993) to compute the power spectrum of the 1999 and 2000 light
curves, after normalizing them to their mean. We used a 3-day binning
(i.e., equal to the typical sampling rate of the source) for both
light curves. As a result, the light curves that we used are evenly
sampled, but there are a few missing points in them; 30 out of 140 and
15 out of 120 for the 1999 and 2000 light curves, respectively. These
points are randomly distributed over the whole light curve, and we
accounted for them using linear interpolation between the two bins
adjacent to the gaps, adding appropriate Poisson noise.

Figure~\ref{figure:psd} shows the PSD of the 1999 and 2000 light
curves (filled circles, and open diamonds, respectively). Each point
in the estimated PSD represents the average of 10 logarithmic
periodogram estimates. Both PSDs follow a power-law like form, with no
obvious frequency breaks. We fitted both spectra with  a
simple power law. The model provides a good
description of both PSDs. The best fitting slope values are consistent
within the errors with the value of $-2\pm 0.3$ in both cases. The
best fitting power law models are also shown in Fig~\ref{figure:psd}
(solid and dashed lines, for the 2000 and 1999 PSDs, respectively).

A closer look at the PSD plot (Fig.~\ref{figure:psd}) indicates a
possible low frequency flattening of the 2000 PSD between
$\log(\nu)=-6.5$ and $-7$, corresponding to a time interval between
$\sim$35 and 115 days.  In order to investigate this issue further we
split the two lowest frequency points into two (to increase the
frequency resolution) and refitted the 2000 PSD with a broken power
law model. We kept the slope below and above the frequency break fixed
to the values $-1$ and $-2$, respectively, and kept as free parameters
the break frequency and normalization of the PSD. The model provides a
good fit to the 2000 PSD, but statistically not better than a simple
power law according to an F-test. The best fitting break frequency is
$2.5\times10^{-7}$ Hz, which corresponds to a break time scale of
$\sim 46$ days. The $90\%$-confidence lower limit to this time scale
is 33 days, while, due to the limited coverage of the observed PSD at
lower frequencies, the upper limit is effectively equal to the length
of the 2000 light curve (i.e. $\sim 300$ days). We also tried to fit
the 1999 PSD with this model, but without success: the best fitting
break frequency turned out to be lower than the lowest frequency
sampled by the 1999 light curve, indicating that there is no
indication of any breaks in the 1999 PSD. This power spectrum follows
a $-2$ power law shape down to the lowest sampled frequencies.

In order to quantify the comparison between the 1999 and 2000 PSDs in
a statistical way, we used the ``{\it S}'' statistic as defined by
Papadakis \& Lawrence (1995). To avoid the constant Poisson noise
level, we considered the periodogram estimates up to frequencies $\sim
7\times 10^{-7}$ Hz. We find that $S=-1.3$, which implies that the two
light curves are not far away from the hypothesis of stationarity.
Therefore, the differences suggested in the previous paragraph by
the broken power-law model fitting results are not confirmed by this
more rigorous statistical test.  Light curves with denser sampling,
which would result in a much higher frequency resolution at lower
frequencies (where the potential differences between the two PSDs are
observed), would be necessary for a more sensitive comparison between
the 1999 and 2000 power spectra.  

To extend the power spectrum estimation to lower frequencies, and
increase the frequency resolution, we then estimated the power
spectrum of the combined 1999 and 2000 light curves. This combined PSD
does not show any frequency break either. It is well fitted by a power
law function, with a slope of $\sim -2$. This result is consistent
with recent results from the power spectrum analysis of combined \xmm\
and \rxte\ light curves of radio-quiet AGN (Markowitz et al., 2003;
McHardy et al. 2004).  According to those results, the $2-10$
keV power spectrum of AGN follows a $-2$ power law form down to a
``break'' frequency, below which it flattens to a slope of $-1$. This
break time scale appears to scale with the black hole mass according
to the relation: $t_{br}(\rm days)=M_{\rm BH}/10^{6.5}~{\rm
M}_{\odot}$ (Markowitz et al. 2003). Using the reverberation
mapping technique, Peterson et al. (2004)
estimate the black hole mass of 3C~390.3 to be $M_{\rm BH}\sim
2.9\times 10^{8}$ M$_{\odot}$.  In this case, the characteristic
break time scale of \object{3C~390.3} should be $\sim 95$ days, which
corresponds to a break frequency of $\sim 10^{-7}$ Hz.  This
break frequency is entirely consistent with the 
broken power law model fitting results in the case of the 2000 light
curve. Unfortunately, though, even the combined 1999 and 2000 PSD does not
extend to frequencies low enough for this break frequency to be 
firmly detected. Longer and/or denser monitoring observations
of this source are needed for the detection of this PSD feature.  
It is worth noticing that the value assumed for the mass of the black
hole harbored by \object{3C~390.3} is still a matter of debate. The
value used in this work is the most recent and it is consistent within
the errors with the mass estimate given by Kaspi et al. (2000).
However, based on another reverberation mapping experiment on this
source, Sergeev et al. (2002) find a longer reverberation lag and
determine a higher mass of $2\times10^9$ M$_{\odot}$ using a different
method. In that case, the putative break frequency in the PSD would be
located at a frequency of $\sim$ 1/yr, which could not be probed with
the current data set.

We conclude that the average, $2-20$ keV PSD of 3C~390.3 follows a
power-law of the form form $P(f)\propto f^{-2}$
at all sampled time scales from $3^{-1}-100^{-1}$
days$^{-1}$, although the presence of a break cannot be excluded
for the 2000 PSD.
The 1999 and 2000 PSDs are statistically consistent with
each other, suggesting that the X-ray emitting process of the source
is ``weakly'' (i.e., second-order) stationary.

\subsection{Analysis in the Time Domain: Structure Function}
\begin{figure}
\psfig{figure=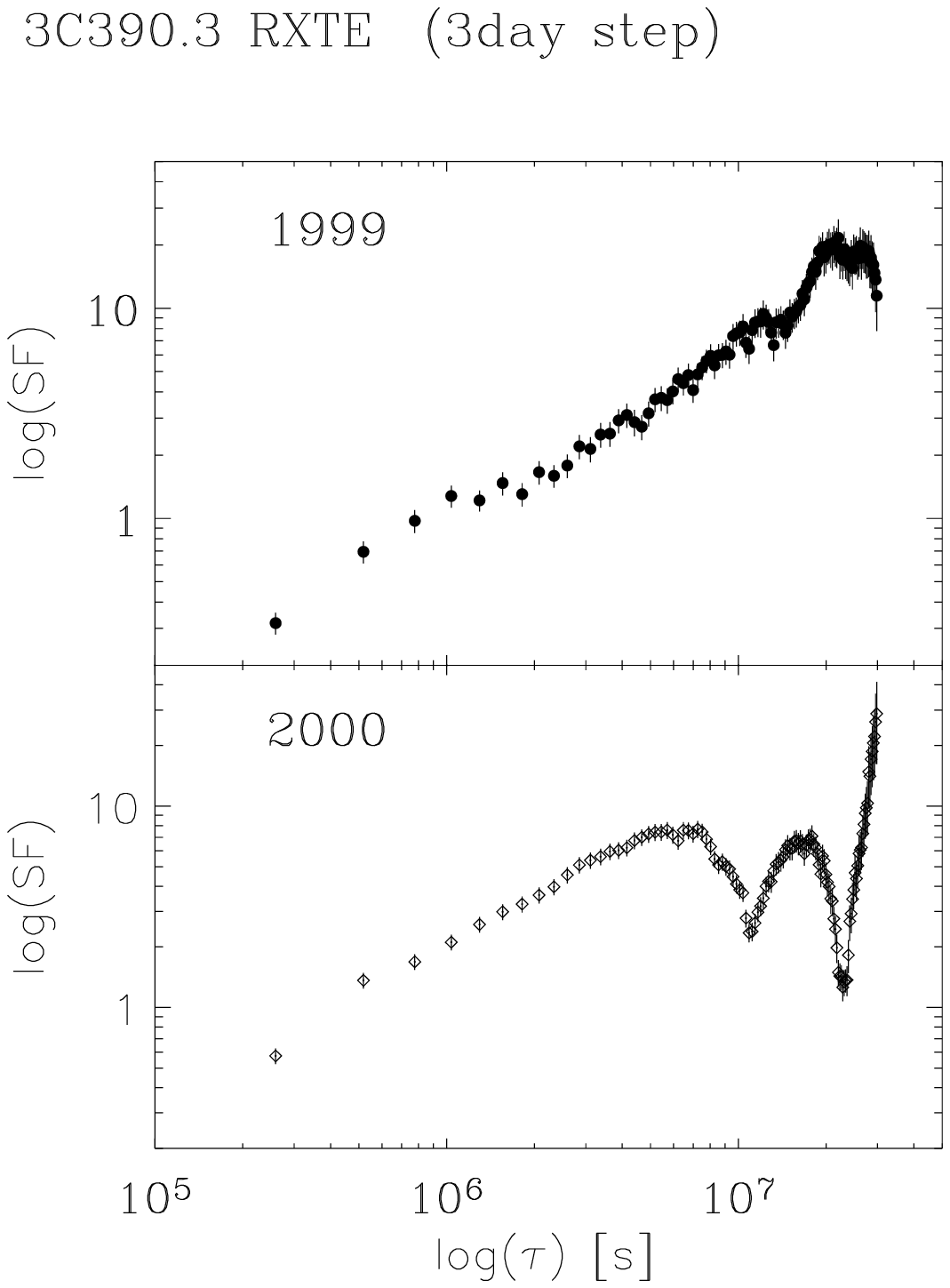,height=10.8cm,width=8.7cm,%
bbllx=110pt,bblly=50pt,bburx=435pt,bbury=435pt,angle=0,clip=}
\caption{Structure function of \object{3C~390.3} during
the 1999 (top panel) and 2000 (bottom panel) monitoring
campaign. Time bins are 3 days.}
\label{figure:SF}
\end{figure}

According to the results from the Fourier analysis presented in Section 4.2,
the ``similarity'' between the 1999 and 2000 PSDs suggests that the
auto-correlation function does not change with time, i.e., the process is
second-order stationary. 
On the other hand, the scaling index method results suggest
significant differences in the statistical properties of the 1999 and
2000 light curves, i.e., the X-ray light curves of 3C~390.3 are
non-stationary. Although this method shows clearly that the two light
curves are intrinsically different, it does not identify what these
differences are.  This apparent difference can be explained in
different ways: (a) the non-stationarity is related to variations of
higher moments than those of second order,
(b) the non-stationarity is related to the non-linear nature of the
3C~390.3 variability (see Leighly \& O'Brien 1997) and linear methods,
such as the PSD, are unable to detect it, or (c) the PSD is not
sensitive enough, due to its intrinsic noise, which requires an
average over many data segments, reducing the length of the time
interval probed.

In order to verify whether the X-ray light curve of 3C~390.3 is indeed
second-order stationary, we performed an analysis based on the
structure function (e.g., Simonetti 1985), a linear method which works
in the time domain and has the ability to discern the range of time
scales that contribute to the variations in the data set.  In
principle, the SF should provide the same information as the PSD, as
both functions are related to the auto-covariance function of the
process $R(\tau)$. The power spectrum is defined as the Fourier
transform of $R(\tau)$, while the intrinsic SF at lag $\tau$ is equal
to $2\sigma^{2}-R(\tau)$, where $\sigma^{2}$ is the variance of the
process (Simonetti et al., 1985). In fact, if the PSD follows a power
law of the form $P(f)\propto f^{-a}$ (where $f$ is the
frequency), then SF $\propto \tau^{a-1}$ (Bregman et al., 1990).

We computed the structure functions for the first and second
monitoring campaigns. The results are plotted in Fig~\ref{figure:SF}:
the top panel (filled circles) refers to the 1999 light curve,
whereas the bottom panel (open diamonds) describes the 2000 SF. Both
SFs have a power law shape of the form SF $\propto \tau^1$, which is
consistent with the results of the PSD analysis, since $P(f)\propto
f^{-2}$ (see \S{4.2}). Furthermore, the 2000 SF shows a flattening to
SF $\propto \tau^0$, at a characteristic time lag of $\sim 4-8\times
10^{6}$ s, followed by an oscillating behavior. This value corresponds
to a characteristic time scale of $\sim 50-200$ days, which
corresponds approximately to the value expected on the basis of the black hole mass
estimate (see discussion on \S{3.1}). The SF flattening cannot be attributed to
the sampling pattern of the 2000 light curve for the following
reasons: 1) the 2000 SF is quite similar to the one derived by Leighly
\& O'Brien (1997) based on a long monitoring campaign carried out by
the \rosat\ HRI, with different duration and sampling; 2) the 2000 SF
is indistinguishable from the one derived using evenly sampled data,
obtained by interpolating linearly the 2000 light curve and removing
any gap. In contrast with the 2000 SF, the 1999 SF does not show a
slope flattening at a similar time lag (the flattening at the highest
time lags, around $2\times 10^{7}$~s, is expected; it is a property of
the SF to approach a value of $\sim 2\times \sigma^{2}$ at the longest
time lags).

The 1999 and 2000 SFs look qualitatively different. However, the
statistical significance of this difference cannot be assessed
directly, since the the SF points are correlated and their true
uncertainty is unknown. The errors shown in Fig.~\ref{figure:SF} are
representative of the typical spread of the points around their mean
in each time-lag bin, hence they depend mainly on the the number of
points that contribute to the SFs estimation at each time lag. To
investigate in a quantitative way whether the difference between the
two SFs can be simply ascribed to the intrinsic, red-noise randomness
of the light curves or it actually reveals a non-stationary behavior,
the use of Monte Carlo simulations is necessary.

One possible way of assessing the significance of the apparent
differences between the 1999 and 2000 SFs, is to perform a
model-independent Monte Carlo experiment.  To this end, we adopted
the method used to assess the significance of the observed delays in
the cross-correlation functions of two simultaneous light curves
(Peterson et al. 1998). This method is based on the ``bootstrap''
technique (Press et al. 1992). Starting from our two observed light
curves, we created two sets of $10^4$ synthetic light curves, by
adding a random offset to each point consistent with its error bar and
then choosing points randomly from the light curve but with the
correct temporal order. Typically, a synthetic light curve is similar
by construction to the original one, but contains 30\% less data
points. Considering the synthetic light curves as multiple
realizations of the processes that produced the two original light
curves, the statistical significance of the difference between the
1999 and 2000 SFs (and therefore the non-stationarity of the light
curve) can be assessed in the following way.

For each pair of synthetic light curves, the SF is calculated and a
parameter quantifying the difference is computed. This quantity is the
sum of the squares of the differences between the two SFs at each time
lag, $D_{\rm SF}\equiv\sum[SF_1(\tau_i)-SF_2(\tau_i)]^2$.  The method
assumes that the differences between the SFs in the $10^{4}$ simulated
pairs, $D_{\rm SF,sim}$, will be distributed around the observed
difference, $D_{\rm SF,obs}$, in a way similar to the distribution of
the $D_{\rm SF,obs}$ values around true value, $D_{\rm SF,true}$. If
the differences in the observed SFs are the result of the red noise
character of the light curves, we would expect that $D_{\rm
  SF,true}=0$. Based now on the distribution of the $D_{\rm SF,sim}$
values, we found that the probability that the $D_{\rm SF,obs}$ value
would be as large as it is, under the assumption that $D_{\rm
  SF,true}=0$, is less than 0.1\%. We conclude that the difference
between the 1999 and 2000 SF is significant at a 99.9\% level.

The model-independent Monte Carlo experiment on the SF indicates 
that the X-ray emitting process
in 3C~390.3 is not even a second-order stationary process, with
the main difference between the 1999 and 2000 light curves arising
at longer time scales.
Up to time
scales of the order of $\sim 100$ days, the 1999 and 2000 SFs resemble
the structure function of red-noise process which has a power law PSD
with slope $\sim -2$, consistent with the PSD results (section
3.1). At lower frequencies the PSD cannot be computed accurately
because the light curves include less than three cycles of any
variability pattern. Although the differences between the
1999 and 2000 PSDs are not statistically significant, the two SFs differ
significantly from each other on long time scales.

\begin{figure}
\psfig{figure=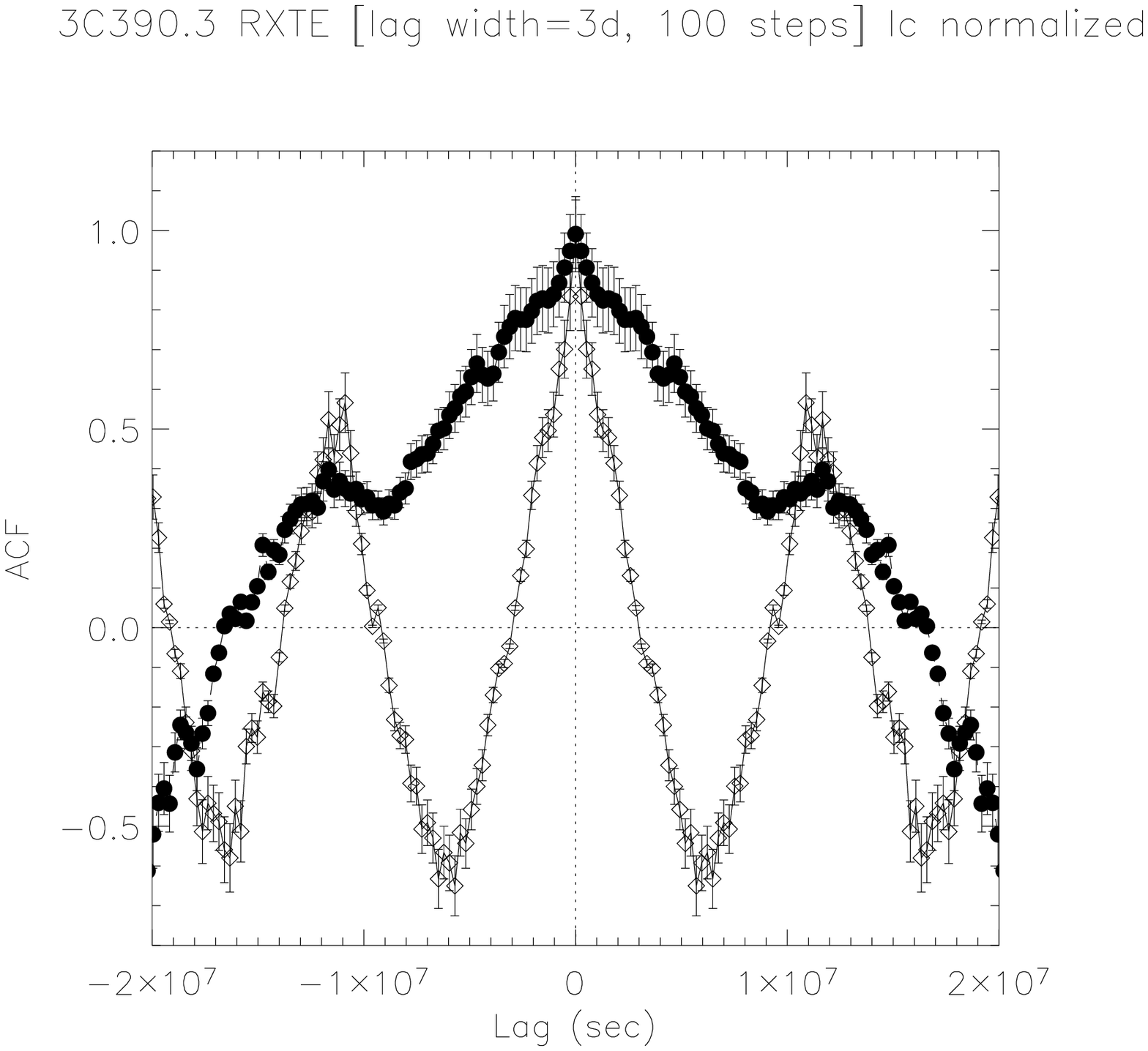,height=8cm,width=8.7cm,%
bbllx=13pt,bblly=5pt,bburx=510pt,bbury=433pt,angle=0,clip=}
\caption{Autocorrelation function of the \object{3C~390.3} light
  curves during the 1999 (filled circles) and 2000 (open diamonds)
  campaigns.  Time bins are 3 days wide.}
\label{figure:ACF}
\end{figure}

The SF results suggest that the
power-law index of the 2000 PSD probably changes from $-2$ to $-1$ at
$f_{\rm br}\sim 1/100$ days, while this break frequency is much lower
during the 1999 observations. This suggestion is entirely consistent
with the PSD fitting results: there is a possibility of a break in
the 2000 PSD, but not in the 1999 PSD.
This ``break-frequency'' variation could
be the reason for the the detection of non-stationarity in the 2 light
curves.

In order to investigate this further, we performed also a
model-dependent numerical experiment, assuming a given
PSD function. The observed 1999 and 2000 PSDs (section 3.1) appear
quite similar, but they extend to frequencies $\sim 2\times 10^{-7}$
Hz only, due to the binning necessary for a reliable PSD estimation.
These frequencies correspond to a time scale of $\sim 5\times 10^{6}$
s, i.e. the time scale above which the most obvious differences in the
observed SFs are evident (Fig.~4). As all model-dependent methods, 
the outcome of these Monte Carlo simulations based on an assumed ``model-PSD'' 
will depend on the assumptions made. 

We assumed that both the 1999 and 2000 light
curves are a realization of a process with a power spectrum of
power-law index $-2$ down to a frequency of 1/(100 days), below which
the power-law index changes to $-1$. The break time scale of 100 days
was found using the Markowitz et al. (2003) relation (see section
3.1), and assuming as estimate of the BH mass for 3C~390.3 the value
obtained by Peterson et al. (2004) from the reverberation mapping.  Then,
we created two sets of 5$\times10^3$ synthetic light curves assuming
this power spectrum model (using the method of Timmer \& Koenig 1995).
The length of the synthetic light curves was 20 times
longer than the length of the observed light curves, and the
points in them were separated by an interval of 3 days. In this way, we
could simulate the effects of the red noise leakage in the estimation of
the SF, although the aliasing effects were not accounted for. This
should not be a serious problem though, as the steep power spectral slope
(-2) implies that these effects should not be very strong. 
Each synthetic light curve was then re-sampled in such a way so that
they have the same length and sampling pattern as the observed light
curves.
The light curves of the first (second) set had the same sampling
pattern, mean, and variance as the 1999 (2000) light curve. The SF of
each simulated light curve was computed, producing a ``synthetic
model'' of the 1999 and 2000 SF. Finally, the observed 1999 and 2000
SFs were compared with these synthetic model.  The results indicate that the
2000 light curve is consistent (at more than 10\% confidence level)
with the ``synthetic model'', i.e., with the hypothesis that the power
spectrum has a break at a time scale of approximately 100~days, while
the 1999 light curve is not consistent with this model
at the 99\% confidence level.  

This result suggests that the PSD break moves to lower frequencies. 
Indeed, when we repeated the numerical experiment with a break 1/400 days, 
the 1999 SF was now consistent with the data (with a probability larger than 
10\%), while the 2000 SF was not (at the 99\% confidence level).
However, if a PSD model with breaks at intermediate frequencies
(e.g., 1/200 -- 1/300 days) is considered, no conclusive results can be drawn,
since the synthetic SF is now marginally consistent
with both the 1999 and 2000 SFs.

Based on the SF analysis and extensive simulations, we conclude that
there is suggestive evidence that one of the differences between the 1999 and 2000 light curves is caused by a change in the characteristic time scale of the system,
although we cannot firmly rule out that the 2 SFs could result 
from the same red noise process (with a PSD showing a break at $\sim$ 1/200-1/300 days).

This conclusion (i.e., the detection of an intrinsic difference in the
temporal properties of the two light curves) is supported by the the
auto-correlation functions of the 1999 and 2000 light curves (see
Fig~\ref{figure:ACF}), estimated by using the Discrete Correlation
Function (DCF) method of Edelson \& Krolik (1988).  The rate at which
the auto-correlation function decays to zero may be interpreted as a
measure of the ``memory'' of the process and thus Fig~\ref{figure:ACF}
indicates that the ``memory'' of the 2000 light curve is shorter than
that of the first 1999 light curve.

\subsection{Time-independent Analysis: Probability Density Function}

In the previous two subsections, we found evidence for
non-stationarity in the light curve of 3C~390.3, by utilizing methods
which are based on the temporal order of the data points.  For the SF
analysis, the dependence on the temporal position of the data points
is trivial, given the definition $SF(\tau)=\langle
[r(t)-r(t+\tau)]^2\rangle$, where $r(t)$ and $r(t+\tau)$
are the light curve data points at times $t$ and $t+\tau$,
respectively. For the non-linear method, this dependence can be easily
recognized, by keeping in mind that the relative positions of the data
points forming a phase space portrait simply reflect the temporal
separations between data points in the light curve.
\begin{figure}
\psfig{figure=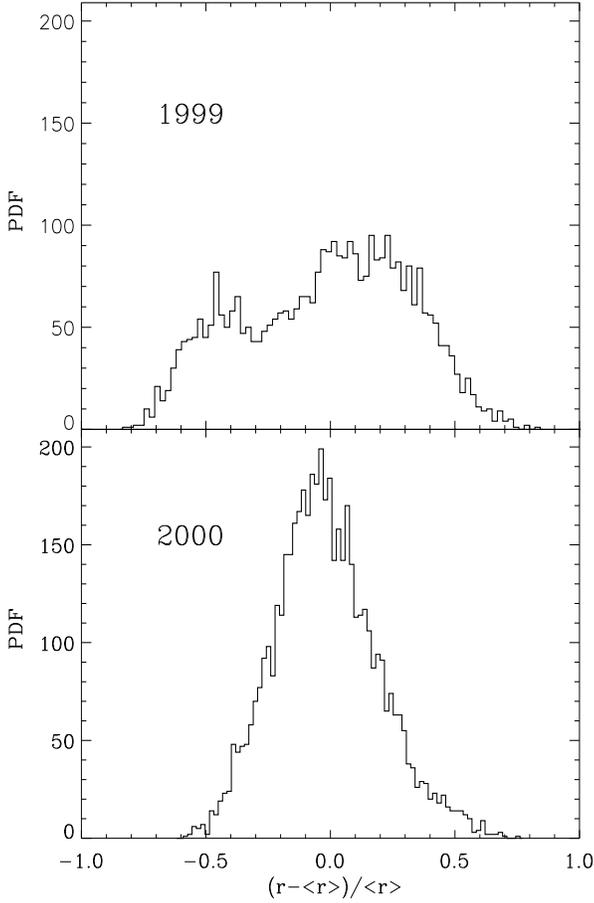,height=12.6cm,width=8.cm,%
bbllx=70pt,bblly=30pt,bburx=400pt,bbury=515pt,angle=0,clip=}
\caption{Probability density function of count rates scaled and
  normalized with respect to the average count rate $\langle r
  \rangle$, during the 1999 (top panel) and 2000 (bottom panel)
  monitoring campaign.  The time bins are 32~s wide.  }
\label{figure:PDF}
\end{figure}

Here we investigate the stationarity of 3C~390.3 with a method
independent of the temporal order of the data points, the probability
density function (PDF), which is the histogram of the different count
rates recorded during the monitoring campaign.  Fig~\ref{figure:PDF}
shows the normalized distributions $(r-\langle r
\rangle)/\langle r \rangle$ of the 1999 (top panel) and 2000
(bottom panel) monitoring campaigns.  The first transformation
($r-\langle r \rangle$) brings both light curves to zero mean,
the second, normalizes the amplitudes to unit mean. This way, the
relative amplitudes of the light curves are directly probed, and the
PDF results can be directly compared to the power spectral ones,
which use exactly the same convention.

For a better visual comparison, we have used small time bins of 32~s
in Fig.~6. However, for a quantitative comparison between PDFs (see
below), we have used larger time bins to prevent any contamination
from the Poisson noise.  There is a clear qualitative difference
between the two PDFs, with the 1999 distribution characterized by two
broad peaks and the 2000 PDF having a bell shape. This
qualitative difference is confirmed by a Kolmogorov-Smirnov (K-S)
test, which gives a probability of 99.99\% that the two distributions
are different.

One may ask what are the effects on the PDF properties of the light
curve bin size, whether the PDF is really time independent (a question
that is directly related to the length of the light curve), and
whether the K-S test is appropriate to quantify the statistical
difference between two PDFs or not. The answers to these questions are
not clear. In all cases, the most important issue is the red-noise
character of the AGN light curves (in all wavebands). As the size of
the light curve bins decreases, the number of the points in the PDF
increases, and the sampled PDF is better defined. However, as the
points added are heavily correlated due to the red-noise character of
the variations, no ``extra'' information is added to the PDF. On the
other hand, since the ``sensitivity'' of the K-S test increases with
the number of points in the sampled PDF (which are assumed to be
independent), there is a possibility that differences which are purely
due to the stochastic nature of a possible stationary process will be
magnified and the K-S test will erroneously indicate ``significant''
differences between two PDFs. Finally, one should also consider the
effects of the Poisson noise to the sampled PDFs. In the cases where
the source signal has a much larger amplitude than the variations
caused by the experimental noise (as is the case here), we should not
expect this effect to be of any significance. In any case, the Poisson
noise should produce similar PDFs (once they are normalized to their
mean) so any significant differences cannot be due to this effect.

To assess the statistical difference between PDFs in a more rigorous
way, we have used two different methods: 1) we constructed PDFs using
small time bins (we used 160 s instead of 32 s to limit the
computational effort) and carried out Monte Carlo simulations in the
same fashion as we did for the SF; 2) we constructed time-independent
PDFs and used the K-S test to assess the significance of the
difference.

In the first approach, we created $10^4$ pairs of synthetic light
curves, utilizing the same model-independent method used for the SF analysis,
and computed their PDFs. We quantified their differences by
computing their squared differences, $D_{\rm
  PDF}\equiv\sum[PDF_1(i)-PDF_2(i)]^2$, and assumed that the
distribution of the $10^{4}$ synthetic $D_{\rm PDF}$ values is
representative of the ``true'' distribution of this quantity, as if we
had actually observed 3C~390.3 for $10^{4}$ times. If the process is
stationary, we would expect that $D_{\rm PDF, real}=0$. Using the
distribution of the $10^{4}$ synthetic $D_{\rm PDF}$ values, we found
that the difference between the 1999 and 2000 PDFs is significant at a
confidence level of more than 99.99\%.

In the second approach, the time independent PDFs were constructed in
the following way. First we binned the 1999 and 2000 light curves at
512 s.  In this way, each observation, which has a typical duration of
$\sim$1000--2000 s, contains up to four contiguous data points. Four
different PDFs were constructed, containing respectively the first,
the second, the third, and the fourth data point of each
observation. This procedure was applied to both light curves, giving
rise to $PDF_{\rm 1a}$, $PDF_{\rm 1b}$, $PDF_{\rm 1c}$, $PDF_{\rm 1d}$
for the 1999 light curve, and $PDF_{\rm 2a}$, $PDF_{\rm 2b}$,
$PDF_{\rm 2c}$, $PDF_{\rm 2d}$ for the 2000 monitoring campaign. In a
way, we can assume that the 8 PDFs are computed from the observations
of 8 individual observers, half of them observing 3C~390.3 in 1999,
and the other half in 2000, each one with a time difference of 512 s
from the other. When we use the K-S test, the four distributions from
the 1999 observations, $PDF_{\rm 1a,b,c,d}$, appear to be identical to
each other.  The same result holds when we compare the four 2000 PDFs,
i.e. $PDF_{\rm 2a,b,c,d}$. This is not surprising, as consecutive
points are strongly correlated, as we mentioned above. However, when
we compare any of the $PDF_{\rm 1a,b,c,d}$ functions with the
respective 2000 function, the K-S test suggests highly significant
differences (with a confidence level higher than $\sim 99.9\%$ for
$PDF_{\rm 1a}$ to be different from $PDF_{\rm 2a}$, and a somewhat
lower confidence level for the other PDF pairs, due to the lower
number of points of their respective distributions).

One possibility for this significant difference between the 1999 and
2000 PDFs is that the one year period is shorter than the
``characteristic'' time scale of the system. As we mentioned in Section
3.1, based on the BH mass estimate of the source, we expect a
characteristic time scale of the order of $\sim 100$ days, shorter
than the duration of the two light curves. However, a second, longer
characteristic time scale must exist.  This time scale will correspond
to the frequency where the PSD should flatten to a slope of $\sim$
zero. This is necessary for the PSD to have a finite value at zero
frequency, otherwise we have to assume an ``infinite'' memory for the
system. Although this time scale is probably longer than the length of
the two light curves, it is very unlikely that this effect can explain
the differences between the 1999 and 2000 PDFs. The reason is that we
have subtracted the respective mean values from the two light curves,
and we have normalized to the mean. In this way, all the long-term
variations, which are not fully sampled by the present light curves,
are suppressed. As a result, most of the variations present in the the
zero-mean, normalized light curves should be caused mainly by
components that are fully sampled.

On the basis of the above results, we can conclude that the PDFs
associated with the 1999 and 2000 light curves are significantly
different, with the former characterized by a bimodal distribution and
the latter by a uniform distribution around the mean, and that the
X-ray emission process in 3C~390.3 is indeed non-stationary.

\begin{figure}
\psfig{figure=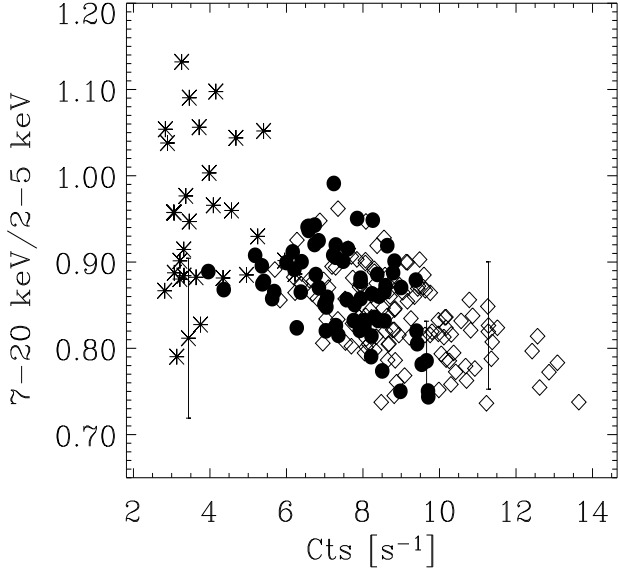,height=8cm,width=8.7cm,%
bbllx=70pt,bblly=60pt,bburx=410pt,bbury=350pt,angle=0,clip=}
\caption{7--20 keV/2--5 keV X-ray color plotted versus the
total count rate of \object{3C~390.3}. Star symbols correspond to the 
first 200 days of the 1999 light curve,
the filled dots describe the second part of the 1999 campaign, and the
open diamond the 2000 light curve. For the sake of clarity, only
a characteristic error for each cluster of points has been plotted. All the
errors associated with the X-ray colors  are
reported in Fig~\ref{figure:lc} bottom panel. 
Time bins are 5760s ($\sim$
1 RXTE orbit).}
\label{figure:HR}
\end{figure}

\section{A Spectral Transition?}
In order to investigate whether the changes of the statistical
timing properties are accompanied  by spectral variations, and specifically
if 3C~390.3 is experiencing an actual state transition, we have performed two
different kinds of analysis.

\noindent
1) The first (model-independent) method is based on the X-ray color -- count rate plot,
often used for GBHs to separate the spectral states.
The plot of X-ray color (i.e., the ratio of the 7--20 over the 2--5 keV count rate)
versus the total 2--20 keV count rate is shown in Fig~\ref{figure:HR}.
The hardness ratio clearly decreases with increasing flux, during the
second part of the 1999 and the whole of the 2000 campaigns. This
trend is typical of type 1 radio-quiet AGN, and has been observed many
times in the past (e.g., Papadakis et al. 2002; Taylor et al. 2004,
and references therein).  This supports the conclusions of Gliozzi et
al. (2002), that the X-ray emission in 3C~390.3 is not jet-dominated.

During the first half of the 1999 campaign, the source was at its
lowest flux state. The spectrum of the source also appears to be quite
hard, consistent with the general trend of spectral hardening towards
low flux states. However, the scatter of the hardness ratio points is
very large, and it seems that during this state, the spectrum
variations are more or less independent of the source flux
variations. So, based on the color--count rate plot, it seems possible
that during this period the X-ray
emission mechanism may behave in a different way, although the errors
associated with the X-ray colors (see Fig~\ref{figure:lc} bottom
panel) prevent us from reaching a firm conclusion.

\begin{table}[ht] 
\caption{Time Intervals Used for Time-Resolved Spectral Analysis}
\begin{center}
\begin{tabular}{ccccc}
\hline
\hline
\noalign{\smallskip}
Start Time  & End Time  & Exposure & Flux${\rm ^a}$ & $\Gamma$ \\
(y/m/d h:m)&(y/m/d h:m)&(ks)& & \\
\noalign{\smallskip}       
\hline
\noalign{\smallskip}
\noalign{\smallskip}
99/01/08 00:26 & 99/02/25 17:58 & 17.90 & 1.87 &$1.55_{0.05}^{0.08}$\\
99/02/28 16:48 & 99/05/11 01:05 & 24.10 & 1.68 &$1.54_{0.05}^{0.03}$\\
99/05/14 00:58 & 99/07/10 20:15 & 21.20 & 2.64 & $1.64_{0.04}^{0.03}$\\
99/07/13 08:56 & 99/09/11 16:03 & 22.56 & 3.86 &$1.68_{0.03}^{0.02}$\\
99/09/14 18:07 & 99/11/13 22:58 & 23.38 & 3.71 &$1.67_{0.02}^{0.02}$ \\
99/11/17 14:25 & 00/01/03 08:00 & 18.69 & 4.71 &$1.74_{0.02}^{0.02}$\\
00/01/06 11:50 & 00/02/29 06:57 & 15.26 & 3.84  &$1.66_{0.03}^{0.02}$\\
00/03/03 04:26 & 00/05/11 12:43 & 32.46 & 4.90 &$1.70_{0.02}^{0.01}$\\
00/05/14 08:41 & 00/07/04 01:22 & 22.50 & 3.80  &$1.61_{0.02}^{0.03}$\\
00/07/07 00:09 & 00/09/01 22:29 & 26.91  & 4.10  &$1.63_{0.02}^{0.02}$\\
00/09/05 5:52 & 00/11/10 12:12 & 32.22  & 3.91   &$1.62_{0.02}^{0.01}$\\
00/11/13 12:57 & 00/12/31 03:12 & 25.47  & 5.03  &$1.67_{0.02}^{0.02}$\\
01/01/03 08:53 & 01/02/24 00:26 & 24.45  & 3.46  &$1.61_{0.02}^{0.02}$\\
\hline
\end{tabular}
\end{center}
${\rm ^a}$ 
2--10 keV flux in units of $(10^{-11}{\rm erg~ cm^{-2} ~s^{-1}})$, calculated assuming a
power law plus Gaussian line model.
\label{table:spec}
\end{table}
\begin{figure}
\psfig{figure=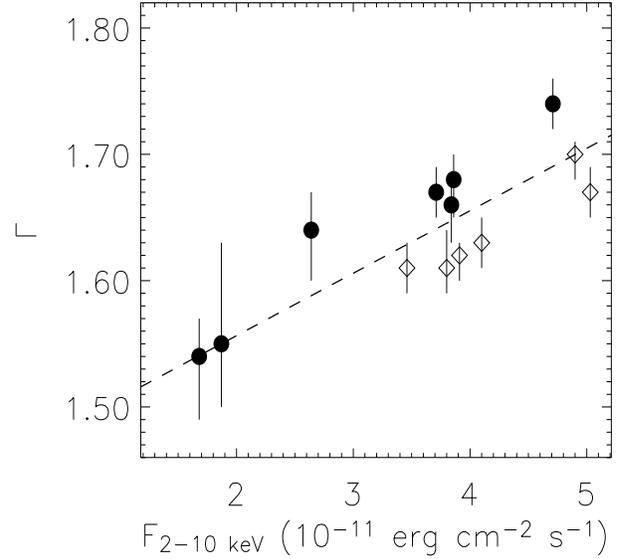,height=8cm,width=8.7cm,%
bbllx=70pt,bblly=60pt,bburx=385pt,bbury=325pt,angle=0,clip=}
\caption{Photon index, $\Gamma$, plotted versus the 2--10 keV flux
($10^{-11}{~\rm erg~cm^{-2}~s^{-1}}$).
Filled dots refers to the 1999 campaign, and the
open diamond to the 2000 light curve. The dashed line represents
a linear least squares fit to all the data points.}
\label{figure:gaflux}
\end{figure}

\noindent 2) More stringent constraints can be derived by using a second (model-dependent) method:
the time-resolved spectral analysis. Since the data consist of short snapshots 
spanning a long temporal baseline, they are well suited for monitoring the 
spectral variability of the source. 

We divided the light curve of Figure~\ref{figure:lc} into thirteen
intervals, of duration 50--60 days each, with net exposures times
ranging between 15 ks and 32 ks. this scheme represents a trade-off
between the necessity to isolate intervals with different but well
defined average count rates and the need to reach a signal-to-noise
ratio (S/N) high enough for a meaningful spectral analysis.  The
dates, exposure times and mean fluxes in the nine selected intervals
are listed in Table~~\ref{table:spec}.

We fitted the 4--20 keV spectra with a model consisting of a power-law, 
absorbed by the Galactic absorbing column, plus a Gaussian \feka\
line with a rest energy fixed at 6.4 keV. In all the intervals, the line is 
statistically required at more than 99\% confidence level, according to an F-test.
However, due to the low S/N, the line parameters are characterized by large
errors ($\sim$ 100\% on the line flux and width) and thus their evolution in time cannot
be investigated properly. On the other hand, the photon indices are well constrained
with uncertainties of the order of 1--3\%. Figure~\ref{figure:gaflux} shows the plot of the
photon index $\Gamma$ versus the 2--10 keV flux, with
the filled dots representing intervals during the 1999 campaign and open diamonds during
the 2000 campaign. The photon index increases steadily with the
2--10 keV flux, without any indication of a discontinuity between the $\Gamma$ corresponding
to the low state and the other photon indices. 
This result, which clearly argues against the
presence of a bona fide state transition, is not totally unexpected if we consider the time
scales at play: in GBHs state transitions take place on time scales of hours-days (e.g.,
Zdziarski et al. 2004), which translate into thousands of years for 3C~390.3, assuming a linear
scaling between the black hole masses.

Based on this analysis, we can conclude that 3C~390.3 is not undergoing a
transition between two distinct spectral states during the \rxte\ monitoring.

\section{Discussion}

We have studied two one-year long, well sampled X-ray light curves of
3C~390.3 from the \rxte\ archive, in search of non-stationarity. This
object is a broad-line radio galaxy, for which a long-term \rosat\
light curve has shown evidence for non-linearity and possibly
non-stationarity in the past at soft X-rays (Leighly \& O'Brien,
1997). The light curves we have used are the longest, and best sampled
light curves on time scales of days/months of this source to
date. Hence, they offer a unique opportunity to investigate the
important issue of stationarity in the X-ray light curves of AGN.

Tests for stationarity of AGN light curves provide important insights
into the physical connection between AGN and GBHs.  In the past few
years, the PSD analysis of long, high signal-to-noise, well sampled
X-ray light curves of a few AGN have shown clearly that the X-ray
variability properties of these objects are very similar to those of
GBHs (e.g. Uttley et al., 2002; Markowitz et al., 2003; McHardy et al., 2004).
However, one
of the most interesting aspects of the X-ray variability behavior of
the GBHs is that they often show transitions to different
``states''. Even when in the same state, their timing properties, like
the PSD, varies with time. For example, during the ``low/hard state'',
the frequency breaks in the X-ray power spectrum of Cyg X-1 change
with time (Belloni \& Hasinger, 1990; Pottschmidt et al., 2003). 
This is one
example of ``stationarity-loss'', which can offer important clues on
the mechanism responsible for the observed variations. ``State''
transitions or, in general, time variations of the statistical
properties of the light curve have not been observed so far in AGN.

\subsection{Multi-technique Timing Approach}
In order to assess whether the light curve of 3C~390.3 is non-stationarity,
we performed a thorough temporal analysis based on different complementary
timing techniques
and extensive Monte Carlo simulations. The first advantage of this approach is that
it leads to more robust results, not biased by the specific
characteristics of a single method. In addition, since the
diverse methods employed probe different statistical properties, they
provide us with complementary pieces of information that can be
combined to constrain the physical mechanism underlying the observed
variability.
\begin{enumerate}
\item The {\it Scaling Index ~method}, with the 1999 distribution peaking at a
value lower than the 2000 one, indicates that the degree of randomness
was higher during the 2000 light curve, which may lead to the
speculation that the number of (coupled) active zones was larger in
2000 than in 1999. However, since the nature (stochastic vs. chaotic)
of the time variability is still unknown, no further physical insights
can be derived from this non-linear tool.  Indeed, as pointed out by
Vio et al. (1992), a naive application of non-linear methods may lead
to conceptual errors, like the detection of low-dimensional dynamics
from signals produced by stochastic processes.  Here the nonlinear
statistics based on the scaling index is used only as a statistical
test to discriminate between two time series, not to determine any
kind of dimension, which would be meaningless for non-deterministic
systems.
 
\item The {\it Power Spectral Density analysis} is the only method
that does not indicate the presence of non-stationarity at a high
significance level.  A difference between the results from PSD and
those from the scaling index method and the PDF may be expected, since
the latter techniques probe higher order statistical properties
compared to the PSD. The difference between the PSD and the SF
analysis may appear more puzzling.  However, since even the PSD
analysis results suggest that the characteristic time scale of the
system moves to lower frequencies in 1999, we believe that both
methods suggest similar trends. The main difference can probably be
ascribed to the necessity for the PSD to average over many data
segments, which reduces the length of the time interval probed, degrades
the frequency resolution, thus preventing the PSD from being a very
sensitive method for detecting non-stationarity.

\item At time scales shorter than $\sim 100$ days the results from the {\it Structure
Function analysis} are consistent with the PSD analysis: the power spectrum of
the source follows a typical power-law form with an index of
approximately $-2$. At longer time scales though, the two SFs are
different: while the 2000 SF appears to flatten to a roughly constant
level (showing several wiggles), the 1999 SF keeps rising up to the
largest time lags sampled. Model-independent Monte Carlo simulations suggest 
that these differences are significant, but no firm conclusions
can be drawn from model-dependent simulations.
On the the other hand, the
periodograms of the two light curves are not significantly different
(as shown by the ``S'' statistic). This result could be due to the low
number of periodogram estimates at these long frequencies. It seems
that the use of the SF analysis can be useful for the detection
of non-stationarity in two light curves, provided that the results
are thoroughly tested with extensive simulations. Although the points in the
sampled SFs are heavily correlated, these functions can still reveal
the presence of significant structure more efficiently than the sampled
power spectrum.

The shape of the 1999 and 2000 SFs at long time scales reveals that
the characteristic time scales are quite different during the two
\rxte\ campaigns. These findings are confirmed by the zero crossing
time scale in the auto-correlation functions (Fig~\ref{figure:ACF}),
showing that the characteristic time scale of the 2000 light curve is
of the order of $\sim$100 d, whereas in the 1999 light curve it should
be at least $2-3$ times longer. 

\item Perhaps, the most telling difference between the timing properties of
the 1999 and 2000 light curves of 3C~390.3 is revealed by the
comparison of the two {\it Probability Density Functions} (see Fig~\ref{figure:PDF}): the data points
in the 1999 light curve follow a bimodal distribution, whereas the
points in the 2000 light curve follow a uniform distribution. This
may imply that the ``anomaly'' causing the non-stationarity occurs
during the first campaign, when two different physical processes seem
to be at work.
\end{enumerate}

The results from the present work, which reveal the presence of
non-stationarity in the X-ray light curves of 3C~390.3, suggest, that
the variability properties of the AGN light curves may also vary with
time, in the same way as they do in GBHs. Therefore, this similarity
further strengthens the analogy between the X-ray variability
properties of AGN and GBHs.
 
\subsection{Analogy between 3C~390.3 and Cyg X-1}
Based on the time-resolved spectral analysis described in $\S 5$, 
we have concluded that 3C~390.3 is not undergoing a genuine
transition between two distinct spectral states.
Nontheless, it is instructive to compare the spectral and temporal behavior
exhibited by 3C~390.3 with the rich phenomenology shown by GBHs on short
time scales within a single spectral state.

Indeed, from the comparison of the the X-ray spectral and temporal 
properties of 3C~390.3 properties with  those of Cyg X--1 in the hard state presented by Pottschmidt et al. (2003), we can infer a qualitative similarity.
In spite of the different methods used to study Cyg X--1 and 3C~390.3,
for both objects we find evidence that both the spectral and temporal 
properties change continuously and in a similar way: when the energy spectrum
becomes harder, the temporal properties are characterized by longer time
scales. 

We can, therefore, conclude that the temporal and spectral behavior
of 3C~390.3, as seen by \rxte\ during the 1999 and 2000 monitoring campaigns, qualitalively mimics the properties of the GBH Cyg X--1 during the low/hard
state, provided that the appropriate time scales are compared. 

\section{Summary and Conclusions}
The main results of this work can be summarized as follows:
\begin{itemize}

\item For the first time, strong evidence of non-stationarity is
  observed in an AGN light curve. This is particularly important in
  view of the analogy hypothesized between AGN and GBHs, since a loss
  of stationary could correspond to a change in the timing properties
  of the X-ray light curves in GBHs. Thus, our results reinforce the
  similarity between the X-ray variability properties of AGN and GBHs.

\item One of the causes of the non-stationarity is likely to be
  the change of the
  characteristic time scale between 1999 and 2000, as suggested by the
  SF and ACF analyses. By analogy with GBH phenomenology, this loss of
  stationarity is likely to correspond to a change in the frequency
  break observed in Cyg X-1 in the low/hard state (e.g., 
  Pottschmidt et al., 2003), rather than a a transition between spectral
  states.

\item The PSD analysis, often considered the most reliable method to
  search for non-stationarity, in this case is probably not sensitive 
  enough to detect it, although there are hints for the presence of a frequency 
  break in the 2000 PSD and not in the 1999 PSD. 
  On the other hand, a number of different timing
  techniques, such as the structure function analysis coupled with 
  Monte Carlo
  simulations, the non-linear scaling index method, and the
  time-independent PDF, prove to be powerful tools to discriminate
between two time series and therefore search for non-stationarity.
  The PDF, in particular, offers a direct demonstration of
  non-stationarity, showing a bimodal distribution in 1999, but a
  uniform distribution in 2000.

\end{itemize}

\begin{acknowledgements}
We thank the anonymous referee for the useful comments and suggestions
that improved the paper
We are grateful to P. Uttley and I.M. McHardy for helpful discussions.
Financial support from NASA LTSA grants NAG5-10708 (MG, RMS),
and NAG5-10817 (ME) is gratefully acknowledged. 
\end{acknowledgements}

\end{document}